\documentclass[11pt]{article}
\usepackage{moriond,epsfig}

\def\Journal#1#2#3#4{{#1} {\bf #2}, #3 (#4)}

\def\PLB{{\em Phys. Lett.}  B}
\def\PRL{\em Phys. Rev. Lett.}
\def\PRD{{\em Phys. Rev.} D}

\def\CQG{\em Class. Quantum Grav.}
\def\RPP{\em Rep. Prog. Phys.}
\def\APJ{\em ApJ}
\def\NAT{\em Nature}
\def\PRT{\em Phys. Rep.}
\def\APP{\em Astropart. Phys.}
\def\MNR{\em MNRAS}
\def\MPL{{\em Mod. Phys. Lett.} A}
\def\PSP{\em Proceedings of the SPIE}
\def\PIA{\em Proceedings of the International Astronomical Union}
\def\PAS{\em Publ. Astron. Soc. Pac.}
\def\PAA{\em Publ. Astron. Soc. of Australia}
\def\AAJ{\em AJ}
\def\NJP{\em New Journal of Physics}
\def\JPC{\em J. Phys. Conf. Ser.}
\def\AAP{\em Astronomy and Astrophysics}

\def\be{\begin{equation}}
\def\ee{\end{equation}}
\def\bea{\begin{eqnarray}}
\def\eea{\end{eqnarray}}

\begin{document}
\vspace*{4cm}
\title{MULTIMESSENGER ASTRONOMY}

\author{N.L. CHRISTENSEN\footnote{nelson.christensen@ligo.org}\\
for the LIGO Scientific Collaboration and the Virgo Collaboration}

\address{Physics and Astronomy, Carleton College,\\
Northfield, Minnesota 55057, USA}

\maketitle\abstracts{Multimessenger astronomy incorporating gravitational radiation is a new and exciting field that will potentially provide significant results and exciting challenges in the near future. With advanced interferometric gravitational wave detectors (LCGT, LIGO, Virgo) we will have the opportunity to investigate sources of gravitational waves that are also expected to be observable through other messengers, such as electromagnetic ($\gamma$-rays, x-rays, optical, radio) and/or neutrino emission. The LIGO-Virgo interferometer network has already been used for multimessenger searches for gravitational radiation that have produced insights on cosmic events. The simultaneous observation of electromagnetic and/or neutrino emission could be important evidence in the first direct detection of gravitational radiation. Knowledge of event time, source sky location, and the expected frequency range of the signal enhances our ability to search for the gravitational radiation signatures with an amplitude closer to the noise floor of the detector. Presented here is a summary of the status of LIGO-Virgo multimessenger detection efforts, along with a discussion of questions that might be resolved using the data from advanced or third generation gravitational wave detector networks.}

\section{Introduction}\label{sec:intro}
The era of gravitational wave (GW) astronomy has begun.
The LIGO~\cite{LIGO} and Virgo~\cite{Virgo} GW interferometric detectors have demonstrated their ability to operate at or near their initial design sensitivities. LIGO's sixth scientific run, S6, and Virgo's third scientific run, VSR3, were recently completed; GEO 600~\cite{GEO} also acquired data during this period.
The LIGO Scientific Collaboration (LSC) and the Virgo Collaboration have been working together in their effort to detect binary inspiral~\cite{LSC-CBC}$^,$~\cite{Virgo-CBC}, burst~\cite{LSC-burst}$^,\,$\cite{Virgo-burst}, continuous wave~\cite{LSC-CW}$^,\,$\cite{Virgo-CW}, and stochastic background~\cite{LSC-Stochastic} signals, as well as GWs associated with electromagnetic (EM) events (such as a $\gamma$-ray burst, GRB)~\cite{LIGO-GRB}$^,\,$\cite{Virgo-GRB}. In 2014, a new generation of detectors, with an even better ability to observe the universe, will come on-line; advanced LIGO (aLIGO)~\cite{aLIGO} and advanced Virgo (AdV)~\cite{aVirgo} will work toward achieving a factor of 10 better sensitivity than the initial detectors. The Large-scale Cryogenic Gravitational wave Telescope (LCGT) is expected to also come on-line around 2015~\cite{LCGT}. A truly global network of advanced detectors will be simultaneously operating in the second half of this decade. The initial ground based laser interferometers were sensitive to GWs in the frequency band from 20 Hz (for Virgo, 40 Hz for LIGO) up to 8 kHz, while the lower frequency for the advanced detectors should drop to 10 Hz.

The existence of GWs was predicted by Einstein \cite{einstein}, and confirmed through observations on the binary pulsar PSR 1913+16. This binary system was discovered in 1974 by Taylor and Hulse~\cite{taylor0}, and subsequent observations by Taylor and Weisberg \cite{taylor} have shown that the decay of the orbit matches perfectly with what is predicted via energy loss by GW emission. 
GW detectors, like LIGO and Virgo, hope to observe GWs produced by astrophysical sources. The observation of these GWs will provide information about the astrophysical event. LIGO, Virgo, and other detectors will not be just GW detectors, they will also be the new generation of astronomical observatories. It is possible that some sources of GWs may not emit EM radiation; for example, imagine the oscillations of a newly formed black hole. Other sources, like a supernova, will likely emit both EM radiation and GWs, and the observation of the GWs in coincidence with EM observations could give new insight about the source. EM observations of the universe are done with radiation having frequencies above 10 MHz. On the other hand, GW observations will be from frequencies below 10 kHz; this should provide very different information about the universe. Since GWs are weakly interacting, any waves produced will traverse the universe without being scattered or absorbed; this gives another unique opportunity for scientists to {\it see} new phenomena in our universe. In this article we discuss how LIGO and Virgo are searching for GW signals in coincidence with EM events. This is an example of multimessenger astronomy. Searches are conducted for GWs at times of observed EM events (the external trigger strategy)~\cite{LIGO-GRB}$^,\,$\cite{Virgo-GRB}. Since GW data from LIGO and Virgo is non-stationary~\cite{S6noise}$^,\,$\cite{VSR2noise}, finding a GW signal candidate in coincidence with an EM transient will increase confidence that the signal is astrophysically produced, and not a spurious noise event.

LIGO and Virgo have developed another strategy for finding GW events in association with EM transients. During a period of joint data collection directional information was sent to EM observatories soon after outlier events were observed in the LIGO-Virgo data; these initial tests took place from Dec 17 2009 to Jan 8 2010, and Sep 4 to Oct 20 2010. When interesting GW triggers were generated, numerous EM observatories have been notified within 30 minutes as part of an EM follow-up effort~\cite{LOOCUP}. 

There are a number of possible sources for an EM signal accompanying a GW.  Long GRBs are likely associated with massive star collapse~\cite{longGRB}, producing $\gamma$-rays then subsequent x-ray and optical afterglows. A double neutron star (NS) or NS/blackhole merger could be the source of short GRBs~\cite{shortGRB} (with prompt $\gamma$-rays and maybe weak, isotropic afterglows). Other interesting phenomena include soft gamma repeater (SGR) flares; these are highly magnetized ($10^{15} G$) neutron stars that emit $\gamma$-ray flares sporadically~\cite{SGR}.

In addition, many astrophysical events will produce detectable high and low energy neutrinos; neutrino events will be another important multimessenger area. LIGO and Virgo are currently working with IceCube~\cite{IceCube}$^,\,$\cite{IceCube2} and ANTARES~\cite{ANTARES}$^,\,$\cite{ANTARES2} in the search for GW signals at the time these neutrino observatories register events. It is suspected that high energy neutrinos could be emitted from long GRBs~\cite{longGRB}, short GRBs~\cite{shortGRB}, low-luminosity GRBs~\cite{lowGRB}, or even {\em choked} GRBs~\cite{chokedGRB}. Core collapse supernovae have prompt low energy neutrino emission (along with delayed optical signals). In the future, with the advanced detectors, it will be fruitful to search for GWs in coincidence with low energy neutrinos from supernovae~\cite{lowneutrino}.

Multimessenger observations could help to address and perhaps resolve a number of open questions in astrophysics~\cite{questions}. For example:\\
{\bf *} What is the speed of GWs? (subluminal or superluminal?)\\
{\bf *} Can GW detectors provide an early warning to EM observers? (to allow the detection of early light curves.)\\
{\bf *} What is the precise origin of SGR flares? (what is the mechanism for GW and EM emission and how are they correlated?)\\
{\bf *} What happens in a core collapse supernova before the light and neutrinos escape?\\
{\bf *} Are there electromagnetically hidden populations of GRBs?\\
{\bf *} What GRB progenitor models can we confirm or reject?\\
{\bf *} Is it possible to construct a competitive Hubble diagram based on GW standard sirens?~\cite{ET}$^,\,$\cite{Nissanke}\\
These are just a few of the astrophysical problems that LIGO and Virgo hope to address with their multimessenger studies.
The remainder of this paper is organized as follows. In Sec.~\ref{sec:results} there is a summary of the multimessenger results to date by LIGO and Virgo. Sec.~\ref{sec:EM} summarizes methods for searching for GW events in coincidence with EM transients, while Sec.~\ref{sec:neutrino} does the same for neutrino events. A conclusion is given in Sec.~\ref{sec:conclu}.

\section{LIGO - Virgo Multimessenger Results}\label{sec:results}
LIGO and Virgo have already published astrophysically important multimessenger papers; while no GWs were observed, the upper limits that have been set
do provide significant constraints on the systems in question~\cite{LIGO-GRB}$^,\,$\cite{Virgo-GRB}$^,\,$\cite{LIGO-GRBa}$^,\,$\cite{LIGO-GRBb}$^,\,$\cite{LIGO-GRBc}. Virgo and LIGO have developed methods whereby searches are conducted for GWs at times of GRBs. By constraining the GW search to a relatively short period (typically tens to hundreds of seconds) the background rejection is improved, and the sensitivity for GW detection is increased. Long GRB events are assumed to be produced by massive star collapse, and GW searches by LIGO and Virgo use their unmodeled {\em burst} search pipelines~\cite{Virgo-GRB}$^,\,$\cite{LIGO-GRBa}$^,\,$\cite{LIGO-GRBb}$^,\,$\cite{LIGO-GRBc}. The coalescence of a neutron star - neutron star, or neutron star - black hole binary system is suspected to be the source of the short GRBs; the LIGO-Virgo {\em compact binary coalescence} and burst pipelines are both used to search for GWs from short GRBs~\cite{LIGO-GRB}. 

Even by not seeing a GW signal in association with a GRB, important astrophysical statements can be made. For example, LIGO and Virgo were able to set lower limits on source distances for 22 short GRBs during LIGO's fifth and Virgo's first scientific runs (S5, VSR1) based on the assumption that these were neutron star - neutron star, or neutron star - black hole binary coalescences~\cite{LIGO-GRB}. For the same S5/VSR1 period, LIGO and Virgo were able to set upper limits on the amplitude of GWs associated with 137 GRBs, and also place lower bounds on the distance to each GRB under the assumption of a fixed energy emission in GWs; the search was conducted for burst waveforms ($< 1s$) with emission at frequencies around 150 Hz, where the LIGO - Virgo detector network had its best sensitivity~\cite{LIGO-GRBc}. The average exclusion distance for the set of GRBs was about 15 Mpc.

The short-duration, hard-spectrum GRB 070201 had an EM determined sky position coincident with the spiral arms of the Andromeda galaxy (M31).
For a short, hard GRB as this was, possible progenitors would be the merger of two neutron stars, a neutron star and a black hole, or a SGR flare. No GW candidates were found in LIGO data within a 180 s long window around the time of this GRB~\cite{LIGO-GRB070201}. The results imply that a compact binary progenitor of GRB 070201 was not located in M31.

SGRs intermittently emit
brief ($\approx 0.1 s$) intense bursts of soft $\gamma$-rays, often with
peak luminosities up to $10^{42} erg/s$; intermediate bursts with greater peak luminosities
can last for seconds. Rare {\em giant flare} events can even be
1000 times brighter than common bursts~\cite{SGR2}. 
SGRs could be good sources of GWs. These {\em magnetars} are likely neutron stars with exceptionally strong magnetic fields (up to $10^{15} G$). The SGR bursts may be from the interaction
of the star’s magnetic field with its solid crust, with
crustal deformations, catastrophic cracking, excitation of the star’s nonradial
modes, and then emission of GWs~\cite{Horvath}. The sources are also potentially close by.
LIGO has conducted searches for short-duration GWs associated with SGR bursts. There was no evidence of GWs associated with any SGR burst in a
sample consisting of the 27 Dec 2004 giant flare from SGR 1806-20~\cite{LIGO-GRBb}, and 190 lesser events from
SGR 1806-20 and SGR 1900+14~\cite{LIGO-SGRa}. An innovative technique was also used to look for repeated GW bursts from the storm of flares from SGR 1900+14; 
the GW signal power around each EM flare was {\em stacked}, and this yielded per burst
energy limits an order of magnitude lower than the
individual flare analysis for the storm events~\cite{LIGO-SGRb}.

\section{Electromagnetic Transients}\label{sec:EM}
There are numerous scenarios where one could expect a GW signal to appear at the same time as an EM event. LIGO and Virgo have recently pursued two strategies to try and find coincident GW and EM events. One is to look for GWs in LIGO and Virgo data at times when EM observatories have registered a transient signal. In the other, LIGO and Virgo have sent times and sky locations to numerous EM observatories with a 30 minute latency; these correspond to LIGO and Virgo {\em triggers} that have been determined to be statistically significant.
\subsection{External Trigger Strategy}\label{subsec:EXT}
Presently there is a search of recent data from LIGO's sixth scientific run (S6) and Virgo's second and third scientific runs (VSR2 and VSR3) for GWs in association with GRBs; LIGO and Virgo are examining events recorded by Swift~\cite{Swift} and Fermi~\cite{Fermi}. Because the time and sky position of the GRB are known, this has the effect of reducing the background noise, and improving the sensitivity of the GW search. 
LIGO and Virgo have also commenced with an effort to find GWs in association with GRBs where the GW signal extends for a time scale of many seconds, to weeks~\cite{STAMP}; the search for these intermediate duration signals has not been previously attempted. 

For long GRBs~\cite{longGRB} LIGO and Virgo use their unmodeled burst pipeline~\cite{LSC-burst}$^,\,$\cite{Virgo-burst} to search for GW signals (since the assumption is that the source is a massive star collapse), while for short GRBs~\cite{shortGRB} they use both the coalescing compact binary search~\cite{LSC-CBC}$^,$~\cite{Virgo-CBC} and unmodeled burst pipelines. The GRBs provide information on the sky position and event time; this simplifies the analysis of the GW data since the time delay between the different GW detectors is known. This also significantly diminishes the data set to be analyzed, reduces the noise background, and therefore increases the sensitivity of the search by about a factor of two~\cite{Was}. For short GRBs a time window for the GW search about the GRB is several seconds; for long GRBs the time window is dictated by GRB astrophysics, and for the LIGO - Virgo search is $-600 s$ to $+ 60 s$.
LIGO and Virgo results for GRB events during S6-VSR2/3 will for forthcoming soon.
\subsection{EM Followups}\label{subsec:LOOCUP}
During two recent periods (17 Dec 2009 to 8 Jan 2010, and 4 Sep 2010 to 20 Oct 2010, within S6-VSR2/3) LIGO and Virgo worked with a number of EM observatories, testing a new method whereby GW data was rapidly analyzed~\cite{LOOCUP}. The time and sky location of statistically significant GW triggers were sent to EM observatories within 30 minutes. Wide EM field of view observations are important to have, but sky location information that is as accurate as possible is also necessary. For this effort the start of the pipeline consisted of triple coincident (from the two LIGO detectors and Virgo) unmodeled burst, or compact binary coalescence triggers. Within a period of 10 minutes it was determined whether the events were statistically significant or not, and whether the quality of the data from the GW observatories was good. The significance above threshold for an event was determined via comparisons with background events. The target false alarm rates were 1 event per day for the initial test period, then reduced to 0.25 event per day for the second test period (excluding Swift~\cite{Swift} and the Palomar Transient Factory~\cite{PTF}, where the rate was 0.1 event per day). Information on known globular cluster and galaxy locations were then used to further restrict the likely sky position of the potential source; only sources out to a distance of 50 Mpc were considered to be possible. Within 30 minutes of the initial registration of the potential GW event, the significant triggers were manually vetted by on-call scientific experts, and scientific monitors in the the observatory control rooms. If a potential GW trigger passed all of the tests the direction information was then sent to various EM observatories, including a number of optical observatories: The Liverpool telescope~\cite{Liverpool}, the Palomar Transient Factory~\cite{PTF}, Pi of the Sky~\cite{Pi}, QUEST~\cite{QUEST}, ROTSE III~\cite{ROTSE}, SkyMapper~\cite{SkyMapper}, TAROT~\cite{TAROT}, and the Zadko Telescope~\cite{Zadko}. Trigger information was also sent to the Swift X-ray observatory~\cite{Swift}$^,\,$\cite{Swift2}, and the radio network LOFAR~\cite{LOFAR}. Part of the research work from LIGO and Virgo has also involved the development of image analysis procedures able to identify the EM counterparts. In the initial S6-VSR2/3 test period there were 8 potential GW events where the information was passed onto the EM observatories, and observations were attempted for 4 of them; for the second test period there were 6 potential GW events, and 4 of them had EM observations attempted.  
The full results from this EM follow-up effort will be published in the near future. This EM follow-up effort during S6-VSR2/3 was a successful milestone, and a positive step toward the advanced detector era where the chances of GW detections will be very enhanced, and these rapid EM observations, when coupled with the GW data, could provide important astrophysical information on the sources.

Long and short GRB afterglows peak a few minutes after the prompt EM/GW emission~\cite{Kann1}$^,\,$\cite{Kann2}, and it is critical to have
EM observations as soon as possible after the GW trigger validation.
{\em Kilo-novae} model afterglows peak about a day after the GW emission~\cite{kilonovae}, so
EM observations a day after the GW trigger would be an important validation for these type of events.
In order to discriminate between the possible EM counterpart (to the GW source) from contaminating transients
repeated observations over several nights are necessary to study the light curve.

\section{Neutrinos}\label{sec:neutrino}
Many of the energetic astrophysical events that could produce GWs are also expected to emit neutrinos. LIGO and Virgo are currently investigating methods to use observations of high and low energy neutrinos to aid in the effort to observe GWs.
\subsection{High Energy Neutrinos}\label{subsec:HEN}
High energy neutrinos (HENs) are predicted to be emitted in astrophysical events that also produce significant amounts of GWs, and by using the time and sky location of observed HENs the ability to confidently identify GWs will be improved. HENs should be emitted in long GRBs; in the prompt and afterglow phases,
HENs ($10^5-10^{10} ~GeV$) are expected
to be produced by accelerated protons in relativistic shocks~\cite{longGRB}. HENs can also be
emitted during binary mergers involving neutron stars~\cite{shortGRB}.
HENs and GWs could both come from low luminosity GRBs; these would be associated with
an energetic population of core-collapse supernovae~\cite{lowGRB}.
There is a class of events where GWs and HENs might be observed in the absence of a GRB observation, namely with choked GRBs; these could plausibly come from
baryon-rich jets. Because the environment could be optically thick, the choked GRB events may be hidden from conventional EM
astronomy, and HENs and GWs will
be the only messengers to reveal their properties~\cite{chokedGRB}.

LIGO and Virgo are presently working with IceCube~\cite{IceCube}$^,\,$\cite{IceCube2} and ANTARES~\cite{ANTARES}$^,\,$\cite{ANTARES2} to see if there are HEN events in coincidence with GW signals in LIGO (S5 and S6) and Virgo (VSR1, VSR2 and VSR3) data. The HEN event time, sky position, and reconstructed energy information enhance the sensitivity of the GW search. During S5 and VSR1 IceCube had 22 of its strings in operation, while ANTARES had 5 strings. IceCube reached its full complement of 86 strings near the time of the end of S6 and VSR3, while ANTARES reached 12 strings. IceCube can provide a neutrino trigger sky location to about 1 degree squared accuracy; then by using catalogs of galaxy positions, including distance, the trigger information from the LIGO and Virgo data can provide a joint test statistic, and reduced false alarm rate. For example, there would be a false alarm rate of about 1 in 435 years for a one-second coincidence time window
and spatial coincidence p-value threshold of 1\%~\cite{HENGW1}$^,\,$\cite{HENGW2}. The size of the time window to be used about the neutrino trigger is a critical parameter in the search, and will need to be larger than 1 s; taking into account the physical processes that could result in neutrino, $\gamma$-ray, and GW emission, it was determined that a conservative $\pm 500 s$ time window would be appropriate~\cite{HENGW3}. The results of this research effort be published soon.

A potential problem for a neutrino - GW search occurs with long GRBs, where HENs from relativistic shocks might be
emitted between a few hours (internal shocks~\cite{lowGRB}) to a few days (external
shocks~\cite{longGRB}) after the GW emission caused by core bounce~\cite{HENGW1}. 
For these events a larger time window will be necessary (days) which will increase the false alarm rate. Better sky position accuracy, either through an improved neutrino detector or an expanded GW detector network (for example with the coming network of advanced detectors), would help to address this issue. 

\subsection{Low Energy Neutrinos}\label{subsec:LEN}
Low energy neutrinos (LENs) will be an important multimessenger partner to GWs for core collapse
supernovae (CCSN). LIGO and Virgo are developing search methods involving LENs, especially for the advanced detector era.
A range of 3 to 5 Mpc is admittedly at the edge of detectability for aLIGO and Super-K~\cite{SuperK}; at this 
distance the supernovae rate becomes about 1/year~\cite{Ando}.
A weak coincident signal in both GWs and LENs may be
convincing, especially if there were also an optical signal.
For a galactic supernova, the neutrino signal will be large,
and LIGO and Virgo would do a standard external trigger search (GRB
search) with a tight coincidence window.
A CCSN produces 10-20 MeV neutrinos (all flavors) over a few 10s of seconds. It is expected that all three neutrino flavors would be created; GWs and neutrinos would be emitted promptly in the CCSN, while EM radiation could be delayed. The neutrino and GW information would truly provide a probe of the physics of the core collapse~\cite{Ott}.
The onset of the signal could probably be determined to better than 1 s. Detectors, such as Super-K~\cite{SuperK}, would detect of order $10^4$ neutrinos for a CCSN at the galactic center. The optical (EM) signature of a CCSN could be obscured; for example, SN 2008iz in M82 was missed via optical observations~\cite{SN2008iz}. With just EM information the exact time of the core collapse bounce could be uncertain to many hours. A tight coincidence window provided by neutrino observations could be used to establish a correlation with GWs. In the advanced GW detector era the sensitivity range of GW and neutrino detectors will be similar, and it is a research goal of LIGO and Virgo that LEN information will be used in association with data from the advanced GW detectors.

\section{Conclusions}\label{sec:conclu}
There is an active effort by LIGO and Virgo to find
GWs in coincidence with EM or
neutrino counterparts. Numerous studies have already been conducted using LIGO and Virgo data from the initial generation of detectors, and more results will be forthcoming soon.
AdV~\cite{aVirgo}, aLIGO~\cite{aLIGO}, and GEO-HF (an upgraded GEO, with improved high frequency response)~\cite{GEO}$^,\,$\cite{GEO-HF} should be on-line in 2014 and start trying to achieve their enhanced sensitivities. LCGT~\cite{LCGT} could be operating in 2015.
A global network of advanced detectors will be simultaneously observing in the second half of this decade, and multimessenger techniques using EM and neutrino event information will improve the probability for detecting GWs. By using GW, EM and neutrino observations all together there will be a tremendous opportunity to decipher the astrophysics pertaining to many different types of cataclysmic events in the universe.

\section*{Acknowledgments}\label{sec:Ack}
This document has been assigned LIGO Document Number P-1100053. The work was funded by NSF Grant PHY-0854790.
The authors gratefully acknowledge the support of the United States
National Science Foundation for the construction and operation of the
LIGO Laboratory, the Science and Technology Facilities Council of the
United Kingdom, the Max-Planck-Society, and the State of
Niedersachsen/Germany for support of the construction and operation of
the GEO600 detector, and the Italian Istituto Nazionale di Fisica
Nucleare and the French Centre National de la Recherche Scientifique
for the construction and operation of the Virgo detector. The authors
also gratefully acknowledge the support of the research by these
agencies and by the Australian Research Council, the Council of
Scientific and Industrial Research of India, the Istituto Nazionale di
Fisica Nucleare of Italy, the Spanish Ministerio de Educaci\'on y
Ciencia, the Conselleria d'Economia Hisenda i Innovaci\'o of the
Govern de les Illes Balears, the Foundation for Fundamental Research
on Matter supported by the Netherlands Organisation for Scientific Research, 
the Polish Ministry of Science and Higher Education, the FOCUS
Programme of Foundation for Polish Science,
the Royal Society, the Scottish Funding Council, the
Scottish Universities Physics Alliance, The National Aeronautics and
Space Administration, the Carnegie Trust, the Leverhulme Trust, the
David and Lucile Packard Foundation, the Research Corporation, and
the Alfred P. Sloan Foundation.


\section*{References}
\begin{thebibliography}{99}
\bibitem{LIGO} B. Abbott {\it et al} (LIGO Scientific Collaboration), \Journal{\RPP}{72}{076901}{2009}.
\bibitem{Virgo} F. Acernese {\it et al} (Virgo Collaboration), \Journal{\CQG}{25}{114045}{2009}.
\bibitem{GEO} H. Grote (LIGO Scientific Collaboration), \Journal{\CQG}{25}{114043}{2008}.  
\bibitem{LSC-CBC} B. Abbott {\it et al} (LIGO Scientific Collaboration), \Journal{\PRD}{80}{047101}{2009}.
\bibitem{Virgo-CBC} F. Acernese {\it et al} (Virgo Collaboration) \Journal{\CQG}{24}{5767}{2007}.
\bibitem{LSC-burst} B. Abbott {\it et al} (LIGO Scientific Collaboration), \Journal{\PRD}{80}{102001}{2009}.
\bibitem{Virgo-burst} F. Acernese {\it et al} (Virgo Collaboration), \Journal{\CQG}{26}{085009}{2009}.
\bibitem{LSC-CW} B. Abbott {\it et al} (LIGO Scientific Collaboration and the Virgo Collaboration), \Journal{\APJ}{713}{671}{2010}.
\bibitem{Virgo-CW} F. Acernese {\it et al} (Virgo Collaboration), \Journal{\CQG}{26}{204002}{2009}.
\bibitem{LSC-Stochastic} B. Abbott {\it et al} (LIGO Scientific Collaboration and the Virgo Collaboration), \Journal{\NAT}{460}{990}{2009}.
\bibitem{LIGO-GRB} J. Abadie {\it et al} (LIGO Scientific Collaboration and the Virgo Collaboration), \Journal{\APJ}{715}{1453}{2010}.
\bibitem{Virgo-GRB} F. Acernese {\it et al} (Virgo Collaboration), \Journal{\CQG}{25}{225001}{2008}.
\bibitem{einstein}  A.~Einstein, Preuss. {\em Akad. Wiss. Berlin}, Sitzungsberichte der physikalisch-mathematischen Klasse, 688 (1916).
\bibitem{taylor0} R.~Hulse and J.~Taylor, \Journal{\APJ}{195}{L51}{1975}.
\bibitem{taylor}  J.~Taylor and J.~Weisberg, \Journal{\APJ}{345}{434}{1989}.
\bibitem{aLIGO} G.M. Harry (LIGO Scientific Collaboration), \Journal{\CQG}{27}{084006}{2010}.
\bibitem{aVirgo} https://wwwcascina.virgo.infn.it/advirgo/docs.html
\bibitem{LCGT} K. Kuroda and the LCGT Collaboration, \Journal{\CQG}{27}{084004}{2010}.
\bibitem{S6noise} N. Christensen (LIGO Scientific Collaboration and the Virgo Collaboration), \Journal{\CQG}{27}{194010}{2010}.
\bibitem{VSR2noise} F. Robinet (LIGO Scientific Collaboration and the Virgo Collaboration), \Journal{\CQG}{27}{194012}{2010}.
\bibitem{LOOCUP} J. Kanner {\it et al}, \Journal{\CQG}{25}{184034}{2008}.
\bibitem{longGRB} E. Waxman and J.N. Bahcall, \Journal{\APJ}{541}{707}{2000}.
\bibitem{shortGRB} E. Nakar, \Journal{\PRT}{442}{166}{2007}.
\bibitem{IceCube} R. Abbasi {\it et al} (IceCube Collaboration), \Journal{\PRD}{83}{012001}{2011}.
\bibitem{IceCube2} F. Halzen and J.P. Rodrigues, \Journal{\CQG}{27}{194003}{2010}.
\bibitem{ANTARES} J.A. Aguilar {\it et al}, \Journal{\PLB}{696}{16}{2011}.
\bibitem{ANTARES2} T. Pradier (Antares Collaboration), \Journal{\CQG}{27}{194004}{2010}.
\bibitem{lowGRB} N. Guptaa and B. Zhang, \Journal{\APP}{27}{386}{2007}.
\bibitem{SGR} K. Ioka, \Journal{\MNR}{327}{639}{2001}.
\bibitem{chokedGRB} S. Horiuchi and Shin'ichiro Ando, \Journal{\PRD}{77}{063007}{2008}.
\bibitem{lowneutrino} C.D. Ott, Adam Burrows, Luc Dessart and Eli Livne, \Journal{\APJ}{685}{1069}{2008}.
\bibitem{questions} S. Marka (LIGO Scientific Collaboration and the Virgo Collaboration), \Journal{\JPC}{243}{012001}{2010}.
\bibitem{ET} B.S. Sathyaprakash {\it et al}, \Journal{\CQG}{27}{215006}{2010}.
\bibitem{Nissanke} S. Nissanke {\it et al}, \Journal{\APJ}{725}{496}{2010}.
\bibitem{LIGO-GRBa} B. Abbott {\it et al} (LIGO Scientific Collaboration), \Journal{\PRD}{72}{042002}{2005}.
\bibitem{LIGO-GRBb} B. Abbott {\it et al} (LIGO Scientific Collaboration), \Journal{\PRD}{77}{062004}{2008}.
\bibitem{LIGO-GRBc} B. Abbott {\it et al} (LIGO Scientific Collaboration and the Virgo Collaboration), \Journal{\APJ}{715}{1438}{2010}.
\bibitem{LIGO-GRB070201} B. Abbott {\it et al} (LIGO Scientific Collaboration), \Journal{\APJ}{681}{1419}{2008}.
\bibitem{SGR2} D.M. Palmer {\it et al}, \Journal{\NAT}{434}{1107}{2005}.
\bibitem{Horvath} J.E. Horvath, \Journal{\MPL}{20}{2799}{2005}.
\bibitem{LIGO-SGRa} B. Abbott {\it et al} (LIGO Scientific Collaboration), \Journal{\PRL}{101}{211102}{2008}.
\bibitem{LIGO-SGRb} B. Abbott {\it et al} (LIGO Scientific Collaboration), \Journal{\APJ}{701}{L68}{2009}.
\bibitem{Swift} http://swift.gsfc.nasa.gov/docs/swift/swiftsc.html
\bibitem{Fermi} http://fermi.gsfc.nasa.gov/
\bibitem{Was} M. Was, PhD thesis, Universit\'e Paris Sud, Document LAL-11-119 (2011).
\bibitem{STAMP} E. Thrane {\it et al}, \Journal{\PRD}{83}{083004}{2011}.
\bibitem{PTF} A. Rau {\it et al}, \Journal{\PAS}{121}{1334}{2009}.
\bibitem{Liverpool} I.A. Steele {\it et al}, \Journal{\PSP}{5489}{679}{2004}.
\bibitem{LOFAR} H.D. Falcke {\it et al}, \Journal{\PIA}{2}{386}{2007}.
\bibitem{Kann1} D.A. Kann {\it et al}, arXiv:0804.1959v2 (2008).
\bibitem{Kann2} D.A. Kann {\it et al},  \Journal{\APJ}{720}{1513}{2010}.
\bibitem{kilonovae} B. D. Metzger {\it et al}  \Journal{\MNR}{406}{2650}{2010}.
\bibitem{Pi} K. Malek {\it et al}, \Journal{\PSP}{7502}{75020D}{2009}.
\bibitem{ROTSE} C.W. Akerlof {\it et al}, \Journal{\PAS}{115}{132}{2003}.
\bibitem{SkyMapper} S. C. Keller {\it et al}, \Journal{\PAA}{24}{1}{2007}.
\bibitem{TAROT} A. Klotz {\it et al}, \Journal{\AAJ}{137}{4100}{2009}.
\bibitem{Zadko} D.M. Coward {\it et al}, \Journal{\PAA}{27}{331}{2010}. 
\bibitem{Swift2} N. Gehrels {\it et al}, \Journal{\NJP}{9}{37}{2007}.
\bibitem{QUEST} C. Baltay {\it et al}, \Journal{\PAS}{119}{1278}{2007}.
\bibitem{HENGW1} Y. Aso {\it et al}, \Journal{\CQG}{25}{114039}{2008}.
\bibitem{HENGW2} E. Chassande-Mottin (LIGO Scientific Collaboration and the Virgo Collaboration) \Journal{\JPC}{243}{012002}{2010}.
\bibitem{HENGW3} B. Baret {\it et al}, \Journal{\APP}{35}{1}{2011}.
\bibitem{SuperK} M. Ikeda {\it et al}, \Journal{\APJ}{669}{519}{2007}.
\bibitem{Ando} S. Ando {\it et al}, \Journal{\PRL}{95}{171101}{2005}.
\bibitem{Ott} C.D. Ott {\it et al}, \Journal{\APJ}{685}{1069}{2008}.
\bibitem{SN2008iz} A. Brunthaler {\it et al}, \Journal{\AAP}{499}{L17}{2009}.
\bibitem{GEO-HF} B. Willke {\it et al}, \Journal{\CQG}{23}{S207}{2006}.





\end{thebibliography}
\end{document}